\documentstyle[11pt,paspconf,psfig]{article}

\def\go{\mathrel{\raise.3ex\hbox{$>$}\mkern-14mu\lower0.6ex\hbox{$\sim$}}}
\def\lo{\mathrel{\raise.3ex\hbox{$<$}\mkern-14mu\lower0.6ex\hbox{$\sim$}}}
\def\fs{\hbox{$.\!\!^{\rm s}$}}
\def\farcs{\hbox{$.\!\!^{\prime\prime}$}}
\def\etal{{\it et~al.\ }}
\def\dm{\,{\rm cm}^{-3}\,{\rm pc}}

\begin{document}
\title{Orbital Parameters of the PSR~B1620$-$26 Triple System }
\author{Z.~Arzoumanian}
\affil{ Department of Astronomy, Cornell University }
\author{K.~Joshi  and  F.~A.~Rasio}
\affil{Department of Physics, Massachusetts Institute of  Technology }
\author{S.~E.~Thorsett }
\affil{Department of Physics, Princeton University }

\begin{abstract}
Previous timing data for PSR~B1620$-$26 were consistent with a second
companion mass $m_2$ anywhere in the range $\sim10^{-3}-1\,M_\odot$,
i.e., from a  Jupiter-type planet to a star. We present the
latest timing parameters for the system, including a significant change in
the projected semi-major axis of the inner binary, a marginal detection of
the fourth time derivative of the pulse frequency, and the pulsar proper
motion (which is in agreement with published values for the proper motion of
M4), and use them to further constrain the mass $m_2$ and the orbital
parameters. Using the observed value of $\stackrel{\ldots.}{f}$,
we obtain a one-parameter family of
solutions, all with $m_2 \lo 10^{-2}\,M_\odot$, i.e., excluding stellar 
masses.
Varying $\stackrel{\ldots.}{f}$ within its formal $1\sigma$ error bar 
does not
affect the mass range significantly. However, if we vary
$\stackrel{\ldots.}{f}$ within a $4\sigma$ error bar, we find that
stellar-mass solutions are still possible.
We also calculate the predicted rate of change of the
projected semi-major axis of the inner binary and show that it agrees with
the measured value.
\end{abstract}


\section{Introduction}

The millisecond radio pulsar PSR~B1620$-$26, in the globular cluster M4, has
a low-mass binary companion (probably a white dwarf of mass
$m_1\approx0.3\,M_\odot$ for a pulsar mass $m_p=1.35\,M_\odot$)
in a $191\,$day low-eccentricity orbit. In addition, its unusually
large frequency second
and third derivatives indicate the presence of a second companion in a wider
orbit around the inner binary (Backer \etal 1993;
Thorsett \etal 1993; Michel 1994).
Such a hierarchical triple configuration is expected to be produced
quite easily in a dense globular cluster through dynamical interactions
between binaries (Rasio \etal 1995; Sigurdsson 1995).

\section{Latest Timing Data }

Timing data were taken at the VLA and Green Bank (140-ft), using
standard hardware and techniques.  Observations at the VLA were made
at 1.6~GHz; observations at Green Bank were made at a variety of
frequencies between 390~MHz and 1.6~GHz.  The data span is 1988 March
to 1995 December.

Timing parameters are reported in Table\,1, and in most cases
represent an incremental improvement over previously published values.
We report for the first time a value for
$\stackrel{\ldots.}{f}$, but emphasize that despite its high
significance in this fit, covariance with unmodeled parameters (e.g.,
$\stackrel{.\ldots.}{f}$) may complicate its interpretation. An
identical fit with a fifth order polynomial yields
$\stackrel{\ldots.}{f}=-1.5\pm2.2\times10^{-40}\mbox{s}^{-5}$ and
$\stackrel{.\ldots.}{f}=-1\pm6\times10^{-48}\mbox{s}^{-6}$ (the other
frequency derivatives are unchanged).

The globular cluster proper motion in RA and Dec is $-9.7(7)$ and
$-12.4(7)$ mas/yr (\cite{ch93b}).  This is about $2\sigma$ from our
timing proper motion, a difference that we do not believe is
significant.

The optical proper motion has magnitude
$\mu=2.42\times10^{-15}\mbox{rad\,s}^{-1}$. This will produce a change
in the projected semi-major axis $x\equiv a_1\sin i$ of $\dot x / x=\mu \sin
j / \tan i,$ where $i$ is the orbital
inclination and $j$ is the angle between the proper motion vector and
the line of nodes  (neither $i$ nor $j$ is known). If the observed
$\dot x$ is due to proper motion, then $\sin j<1$ implies
$i<15^\circ$, which in turn implies that the companion must have mass
$>1.6M_\odot$, i.e., it is a neutron star or black hole, not a white
dwarf.  The nearly circular orbit is then problematic, but could be
explained if the companion were captured in a three or four
body interaction which involved a collision, followed by
circularization during a Thorne-Zytkow phase or by a disk of material
from a disrupted star.  In our opinion, however, it is more likely
that the observed $\dot x$ has, instead, some contribution from
another source, such as precession.

At the most likely orbital inclination $i=60^\circ$, proper motion can
account for only about 10\% of the observed $\dot x$.  A more
attractive explanation is that $\dot x$ is caused by precession of the
binary orbit in the gravitational field of the same third body that
has been invoked to explain the large spin-frequency derivatives. It can be
shown that this ``planetary'' precession
yields $\dot x / x = Gm_2 F / (\Omega_1 r_2^3)$, where 
$\Omega_1$ is the angular velocity of the inner
orbit and $r_2$ is the distance to the second companion
(assumed fixed).  
$F$ is a complex function of unknown angles, but has a median value for
random orientations of $0.12$, a 50\% range of $0.033-0.33$, and a
95\% range of $0.0013-1.6$.

Assuming the inner binary has mass $M=m_p+m_1=1.7\,M_\odot$ and 
a low-eccentricity outer
orbit, and ascribing the entire $\dot x$ to precession, we can write
$F=(P_2 / 3360\mbox{\,yr})^2  (1+ M/m_2),$
where $P_2$ is the period of the outer orbit, and calculate $F$ for
the three illustrative solutions of Rasio (1994). For
$m_2=80_\oplus$ and $P_2=10$\,yr, $F=0.06$, while for $m_2=0.8M_\odot$ and
$P_2=120$\,yr, $F=0.004$, and for $m_2=1.4M_\odot$ and $P_2=500$\,yr,
$F=0.04$. We conclude that the measurement of $\dot x$ is consistent
with the presence of a third companion, and indeed is a strong
indication that the triple hypothesis is correct.  However, by itself
it is only a poor discriminator between stellar and planetary solutions.
With more information from spin frequency derivatives and from other
orbital perturbations such as $\dot\omega$ and $\dot P_b$, the
measurement of $\dot x$ will help untangle the various unknown angles
that characterize the system. A complete analysis of these orbital
perturbation effects will be presented elsewhere (Joshi \& Rasio 1996).

\section{Orbital Parameters }
 
The best test to confirm the triple nature of this system would be a
Keplerian fit to the timing data spanning more than one orbit of the outer
body. However, since the inferred orbital period $P_2$
is of order a century or
more, we observe only a small portion of an orbit, effectively measuring
successively higher order derivatives of the acceleration at a single point.
If derivatives of the pulse frequency up to the {\em fifth order\/}
were available,
the system could in principle be solved completely (except for the usual
inclination angle) for the orbital parameters and mass of the second
companion. Here we use the values listed in Table~1 for frequency
derivatives up to the fourth order to obtain a one-parameter family of
solutions. While computing the orbital parameters of the second companion,
we treat the inner binary as a single object of mass $M = 1.7 M_\odot$.
For fixed values of the eccentricity $e_2$ and the inclination $i_2$, we
solve the non-linear system of equations for the frequency derivatives
(using the Newton-Raphson method) for the mass $m_2$, the semi-major axis
$a_2$, the angle of periastron $\omega_2$ and the longitude from pericenter
$\lambda_2$ (Joshi \& Rasio 1996).

Fig.~1 illustrates our ``standard solution'', obtained using the
observed value of $\stackrel{\ldots.}{f}$. We see that there are no
solutions for $e_2  \lo 0.1$. Hence a nearly circular orbit is ruled out.
For $0.1 \lo e_2 \lo 0.3$ there are two solutions for each value of the
eccentricity, and hence two possible values of $m_2$. In one solution, $m_2$
approaches zero as $e_2$ approaches $\approx0.3$. However for
$m_2 \lo 10^{-5}\,M_\odot$
the triple configuration becomes dynamically unstable.
In addition, we can rule out these
solutions because they have very short
orbital periods ($P_2 \lo 5\,$yr) and this would have been detected already
in the timing data.
For the
other solutions, $m_2$ increases monotonically in the range
$\sim10^{-3}-10^{-2}\,M_\odot$ (Jupiter to brown-dwarf masses)
as $e_2$ increases from $\sim0.1$ to $1$.

In Fig.~2 we show the results obtained on varying $\stackrel{\ldots.}{f}$
within a $4\sigma$ error bar around the best fit value 
$\stackrel{\ldots.}{f}_{m}$ (given in Table~1).
The results are shown for
$\stackrel{\ldots.}{f}/\stackrel{\ldots.}{f}_{m}$ = $1.0$ (as in Fig.~1,
solid line), $2.3$ (long-dashed line), $0.001$ (short-dashed line),
$-0.05$ (dotted line) and $-0.4$ (dot-dashed line).
We find that stellar-mass solutions for $m_2$ are still possible if
$-0.05 \lo \stackrel{\ldots.}{f}/ \stackrel{\ldots.}{f}_{m} \lo 0.001$.
A second companion of stellar mass would provide
a natural explanation for the eccentricity of the inner
binary in terms of secular
perturbations (Rasio 1994) and would also be consistent with a
preliminary identification of an optical counterpart for the system
(Bailyn \etal 1994).

\begin{figure}[t]
{\psfig{figure=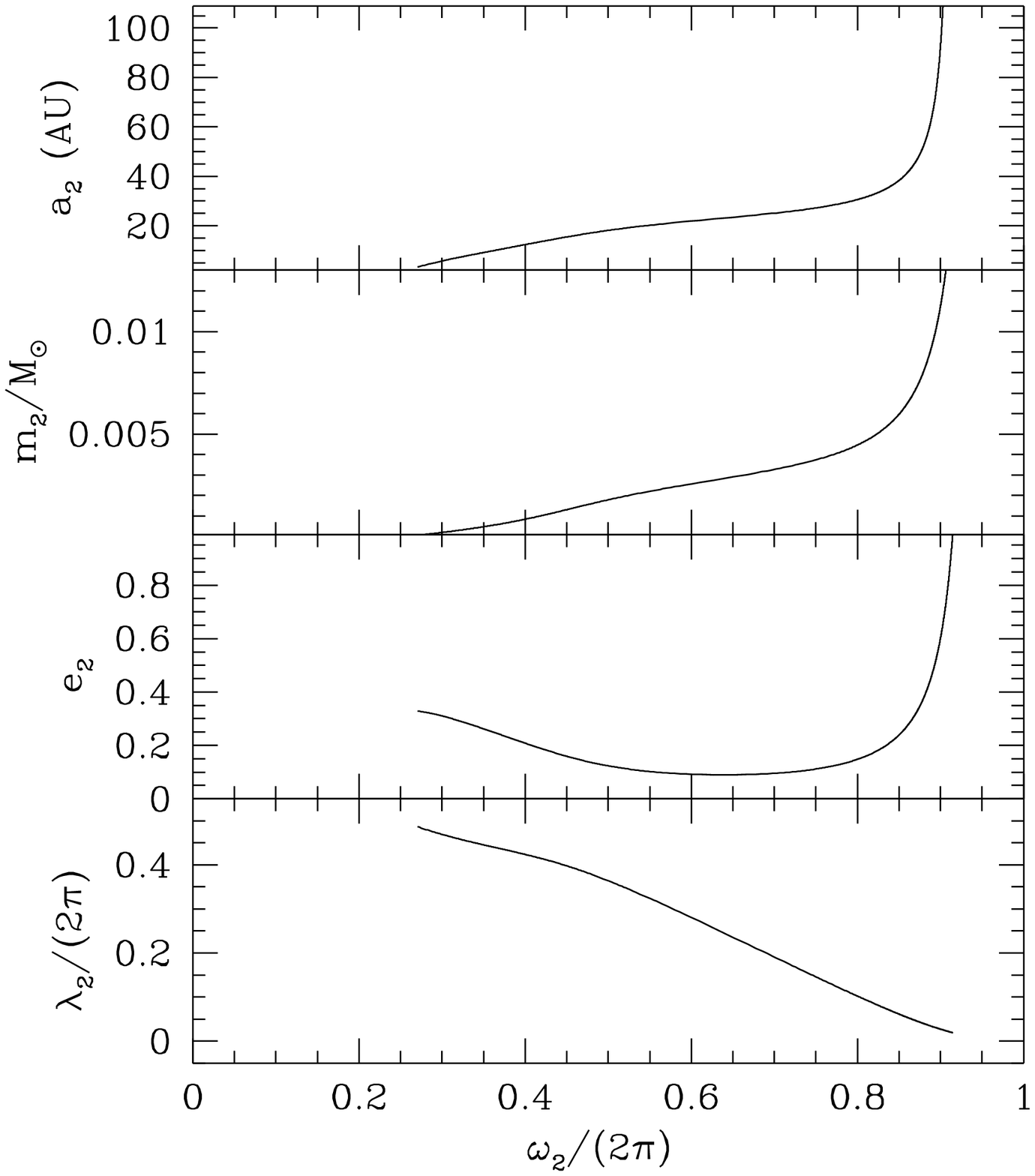,height=5in}}
\caption{Our ``standard solution,'' using the observed value of
$\stackrel{\ldots.}{f}$.}
\end{figure}

\begin{figure}[t]
{\psfig{figure=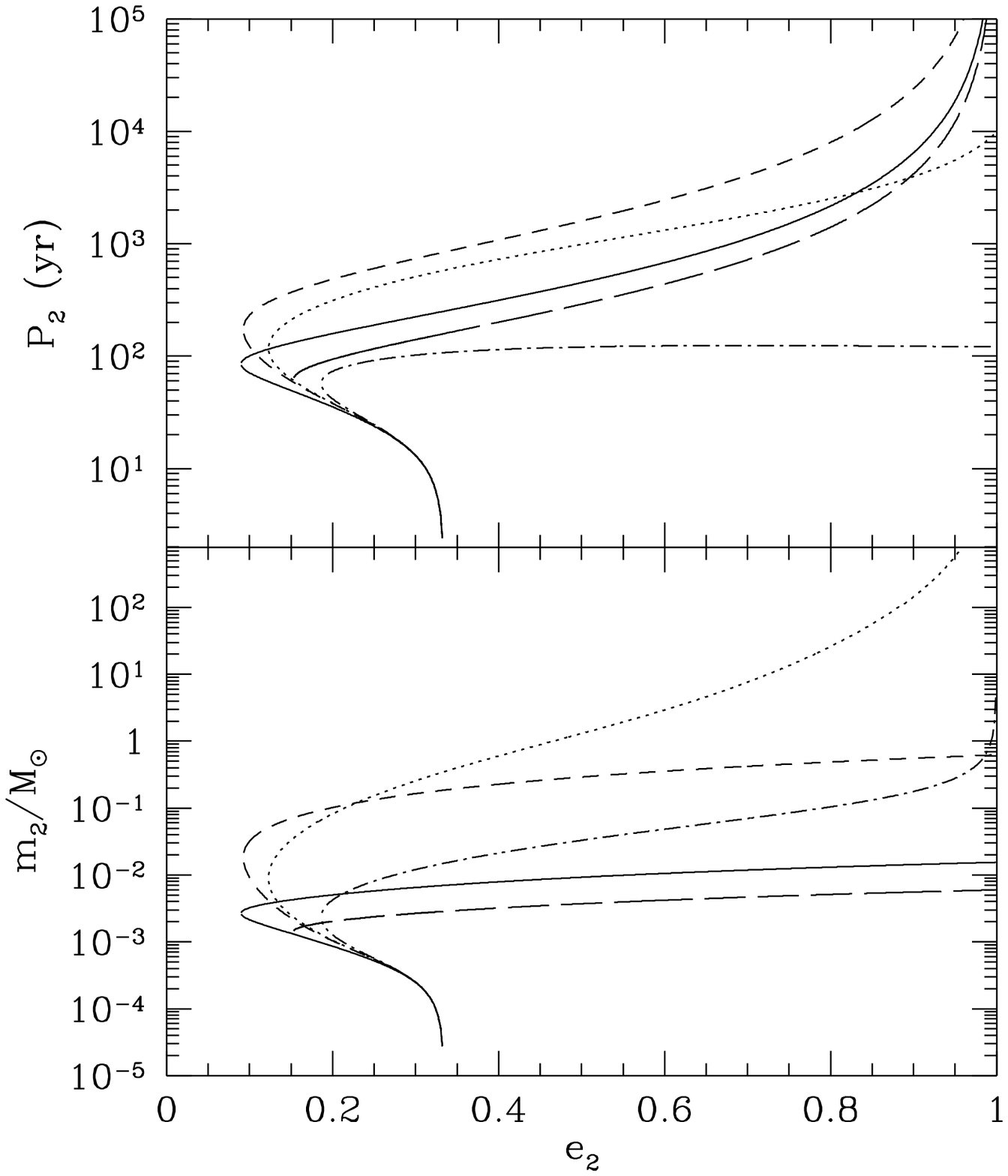,height=5in }}
\caption{The solutions for
$\stackrel{\ldots.}{f} = \stackrel{\ldots.}{f}_{m}$ (solid line),
for $\stackrel{\ldots.}{f} = 2.3 \stackrel{\ldots.}{f}_{m}$
(long-dashed line),
for $\stackrel{\ldots.}{f} = 0.001 \stackrel{\ldots.}{f}_{m}$
(short-dashed line),
for $\stackrel{\ldots.}{f} = -0.05 \stackrel{\ldots.}{f}_{m}$
(dotted line) and
$\stackrel{\ldots.}{f} = -0.4 \stackrel{\ldots.}{f}_{m}$
(dot-dashed line).
We assume here that the inclination angle $i_2=90^\circ$.}
\end{figure}

\begin{table}[t]
\begin{center}
\begin{tabular}{ll}
\hline
Right ascension (J2000.0) & $16^{\mbox{\scriptsize
h}}23^{\mbox{\scriptsize m}}38\fs2228(6)$ \\
Declination (J2000.0) & $-26^\circ31'53\farcs74(4)$ \\
Proper motion RA (mas\,yr$^{-1}$) & $-$16(3) \\
Proper motion Dec(mas\,yr$^{-1}$) & $-$30(18) \\
Dispersion measure ($\dm$) & 62.8627(8) \\
\\
Spin period $P$ (ms) & 11.075750892214(6) \\
Spin frequency $f$ (Hz) & 90.2873321847(5) \\
$\dot f$~(s$^{-2}$) & $-6.065(2)\times10^{-15}$ \\
$\ddot f$~(s$^{-3}$) & $1.897(4)\times10^{-23}$ \\
$\stackrel{\ldots}{f}$ (s$^{-4}$) & $1.2(2)\times10^{-32}$ \\
$\stackrel{\ldots.}{f}$ (s$^{-5}$) & $-1.7(5)\times10^{-40}$ \\
Epoch of $f$ (MJD) & 48365.0 \\
\\
Projected semi-major axis $x$ (s) & 64.809478(8) \\
Orbital period $P_b$ (s) & 16540653(6) \\
Eccentricity $e$ & 0.0253151(3) \\
Time of periastron $T_0$ (MJD) & 48345.3771(3) \\
Angle of periastron $\omega$ & 117.1296(6) \\
Mass function ($M_\odot$) & $7.975\times 10^{-3}$ \\
\\
Advance of periastron $\dot\omega$ ($^\circ\mbox{yr}^{-1}$) &
$(-2.9\pm2.4)\times10^{-4}$ \\
$\dot P_b$ & $(1.7\pm2.6)\times10^{-9}$ \\
$\dot e$ (s$^{-1}$) & $5(4)\times 10^{-15}$ \\
$\dot x$ & $6(1)\times10^{-13}$ \\
\hline
\end{tabular}
\end{center}
\caption{\label{params}Timing parameters of PSR~B1620$-$26. Position is
relative to the JPL~DE202 solar system ephemeris.  Numbers
in parentheses are uncertainties in the final digits quoted. {\em
  NOTE:} formal uncertainties are relative to model fit with above
parameters. Covariances with unfit parameters (e.g., the fifth
frequency derivative) may increase true uncertainties, particularly of
$\stackrel{\ldots}{f}$ and $\stackrel{\ldots.}{f}$.}
\end{table}

\end{document}